\documentclass{PoS}
\usepackage{url}
\usepackage{amsmath}
\usepackage{gensymb}
\usepackage[utf8]{inputenc}

\title{An exploratory study of heavy-light semileptonic form factors using distillation}

\ShortTitle{Heavy-light semileptonic form factors using distillation}

\author{Peter Boyle\\
        Higgs Centre for Theoretical Physics, University of Edinburgh, EH9 3FD, Edinburgh, UK\\
Physics Department, Brookhaven National Laboratory, Upton, NY, USA
\\
        E-mail: \email{paboyle@ed.ac.uk}}

\author{\speaker{Felix Erben}\\
        Higgs Centre for Theoretical Physics, University of Edinburgh, EH9 3FD, Edinburgh, UK\\
        E-mail: \email{felix.erben@ed.ac.uk}}
        
\author{Michael Marshall\\
        Higgs Centre for Theoretical Physics, University of Edinburgh, EH9 3FD, Edinburgh, UK\\
        E-mail: \email{michael.marshall@ed.ac.uk}}
                
\author{Fionn \'O h{\'O}g{\'a}in\\
        Higgs Centre for Theoretical Physics, University of Edinburgh, EH9 3FD, Edinburgh, UK\\
        E-mail: \email{fionn.o.hogain@ed.ac.uk}}

\author{Antonin Portelli\\
        Higgs Centre for Theoretical Physics, University of Edinburgh, EH9 3FD, Edinburgh, UK\\
        E-mail: \email{antonin.portelli@ed.ac.uk}}
        
\author{Justus Tobias Tsang\\
        Higgs Centre for Theoretical Physics, University of Edinburgh, EH9 3FD, Edinburgh, UK\\
        CP$^3$-Origins and IMADA, University of Southern Denmark, Campusvej 55,DK-
5230 Odense M, Denmark\\
        E-mail: \email{tsang@imada.sdu.dk}}        

\abstract{We present our exploratory study with the aim of simulating heavy-light semileptonic form factors as part of the RBC-UKQCD charm (to bottom) physics programme. We are using a distillation-based setup as a strategy to get optimised plateaus in semileptonic $D_{(s)}$ and $B_{(s)}$ decays, and compare our results to form factors obtained from sequential $Z_2$-Wall
propagators. The study is done in a centre-of-mass frame as well as in several moving frames. We use an $N_f=2+1$ domain wall fermion ensemble with a pion mass of $340$ MeV, with the aim of extending the study to a variety of other domain-wall ensembles, including physical-pion mass ensembles.}

\FullConference{37th International Symposium on Lattice Field Theory - Lattice2019\\
		16-22 June 2019\\
		Wuhan, China}

\begin{document}

\section{Introduction}
Flavour physics, i.e. physics studying weak processes which change quark flavour, could give access to potentially new physics beyond the Standard Model. These flavour changes are described by the Cabibbo-Kobayashi-Maskawa (CKM) matrix \cite{Cabibbo:1963yz,Kobayashi:1973fv}. Heavy-light semileptonic processes, like $D \rightarrow \pi \ell \nu$,
give access to its elements ($|V_{cd}|$ in the process mentioned). Currently, lattice QCD results are in agreement with the Standard Model prediction of a unitary CKM matrix, but a further reduction in error could potentially lead to new physics. Another motivation for the study of heavy-light semileptonic processes involving $B$ mesons are the R-ratios
\begin{align}
R(D^{(*)})  = \frac{\mathcal{B}(B \rightarrow D^{(*)} \tau \nu_\tau)}{\mathcal{B}(B \rightarrow D^{(*)} \ell \nu_\ell)}
\, , \qquad
\ell = e,\mu
\, ,
\end{align}
where currently a tension \cite{Amhis:2019ckw} in lepton flavour universality between experiment and theory is observed, giving rise to the need of a clear first-principles determination of these ratios.

\begin{figure}[htbp] % no figure before 1st section
  \centering
\includegraphics[width=0.34\textwidth]{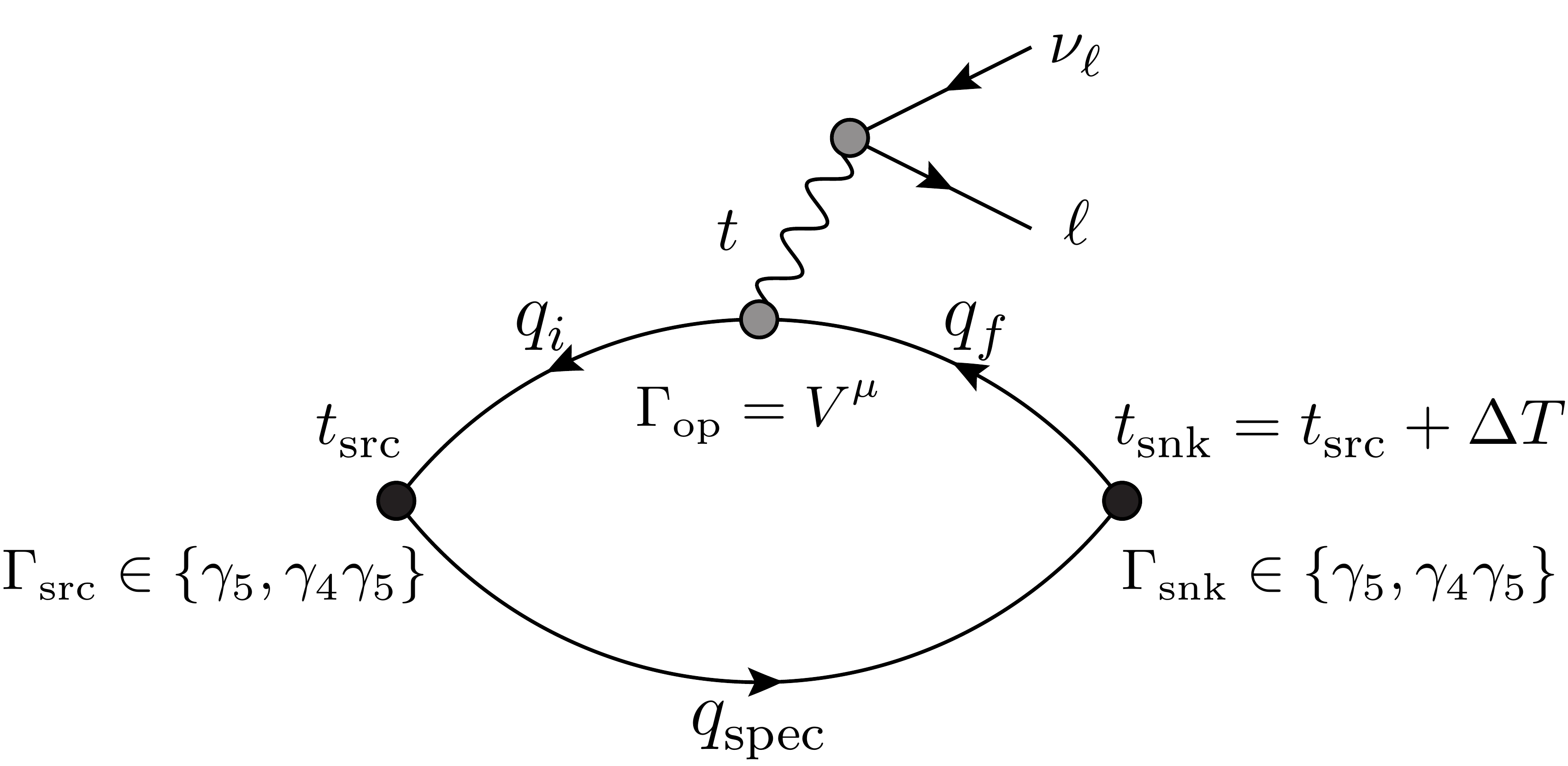}
  \caption{Diagram of the tree-level three-point function of semileptonic decays. For $D \rightarrow \pi \ell \nu$, $q_\mathrm{spec}=l$, $q_\mathrm{i}=h$, $q_\mathrm{f}=l$, where $h$ denotes a heavy quark and $l$ denotes a light quark.}
  \label{figure:hl-semi}% Give a unique label
\end{figure}

The main observable computed on the lattice for the study of heavy-light semileptonic processes are the three-point functions
\begin{align}
C_3(\Delta T = t_\mathrm{snk} - t_\mathrm{src},t) = \langle \Gamma_\mathrm{snk} D_{q_\mathrm{f}}^{-1}(t_\mathrm{snk},t) \Gamma_\mathrm{op} D_{q_\mathrm{i}}^{-1}(t,t_\mathrm{src}) \Gamma_\mathrm{src} D_{q_\mathrm{spec}}^{-1}(t_\mathrm{src},t_\mathrm{snk}) \rangle
\, ,
\end{align}
shown diagrammatically in Figure \ref{figure:hl-semi}.
They suffer from a bad signal-to-noise ratio, rendering their computation a challenging task. Advanced numerical methods and algorithms are needed to tackle the computation of heavy-light semileptonic form factors effectively. We present a feasibility study which uses distillation with LapH smearing \cite{Peardon:2009gh,Morningstar:2011ka} to estimate the relevant correlation functions and compare it to sequential propagator inversion, as used in RBC/UKQCD's heavy-light semi-leptonic form factor calculations using the relativistic heavy quark action \cite{Flynn:2019any}. Both approaches use the highly optimised lattice QCD code Grid \cite{Boyle:2016lbp}, together with Hadrons \cite{Portelli:Hadrons}.

\section{Computation of three-point functions}

The dominating contributions to the weak decay in these three-point functions $C_3$ (Figure \ref{figure:hl-semi}) are short-distance. We therefore can treat this operator as point-like. Experimentally the region around $q^2=0$ is most precisely known. In our lattice QCD calculation, $q^2 = (E_D-E_\pi)^2 - (\mathbf{p}_D - \mathbf{p}_\pi)^2$, with the energies $E_D,E_\pi$ and momenta $\mathbf{p}_D ,\mathbf{p}_\pi$ of the $D$ meson and pion, is larger than zero for small momenta. By choosing different momenta for the two particles, we can map out the $q^2$ region and approach or extrapolate towards $q^2=0$ - because of the much smaller rest mass of the pion, it is beneficial to keep $\mathbf{p}_D=\mathbf{0}$ and to vary $\mathbf{p}_\pi$.

\begin{figure}[htbp] % no figure before 1st section
  \centering
\includegraphics[width=0.45\textwidth]{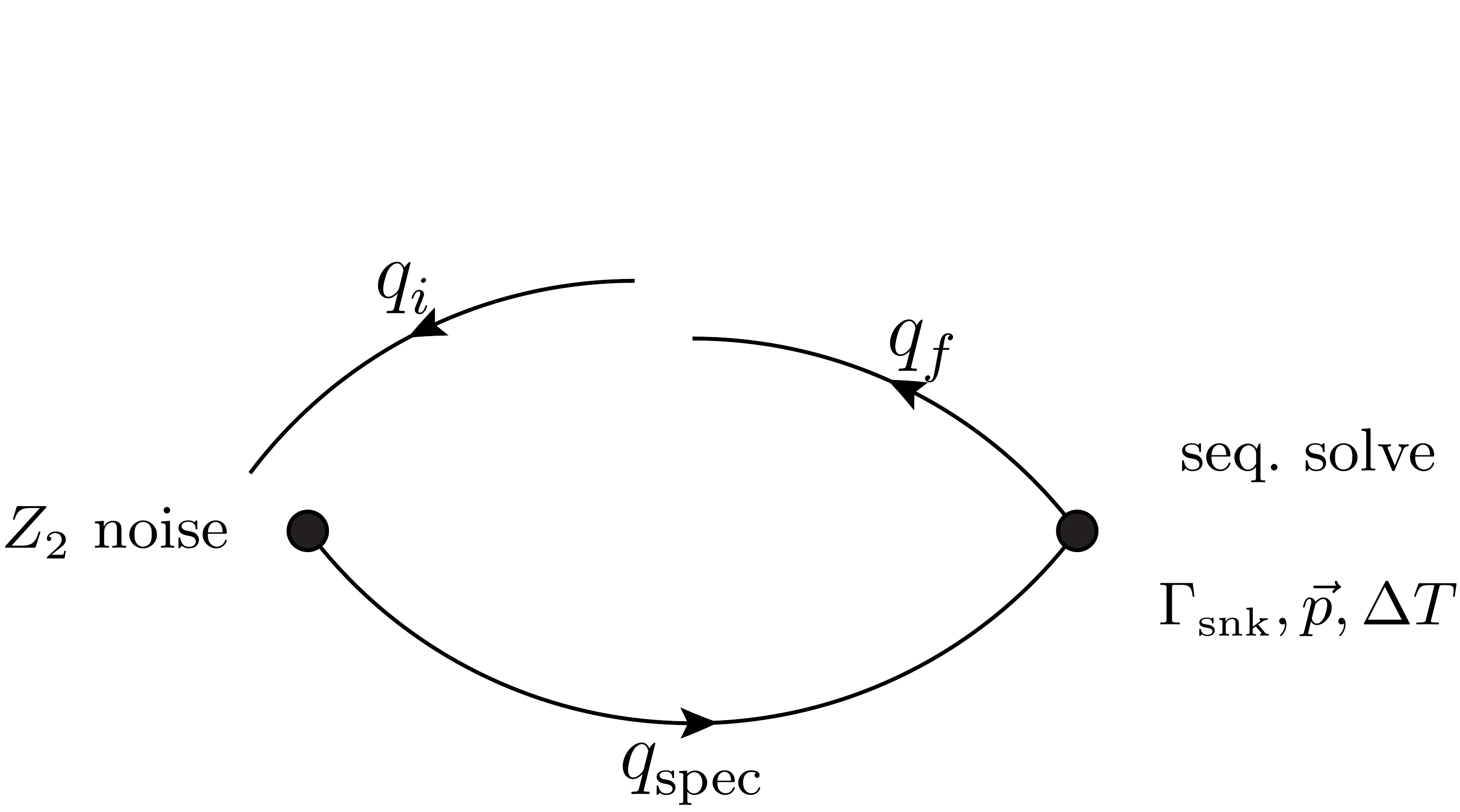} \vline \vspace{5mm} \includegraphics[width=0.35\textwidth]{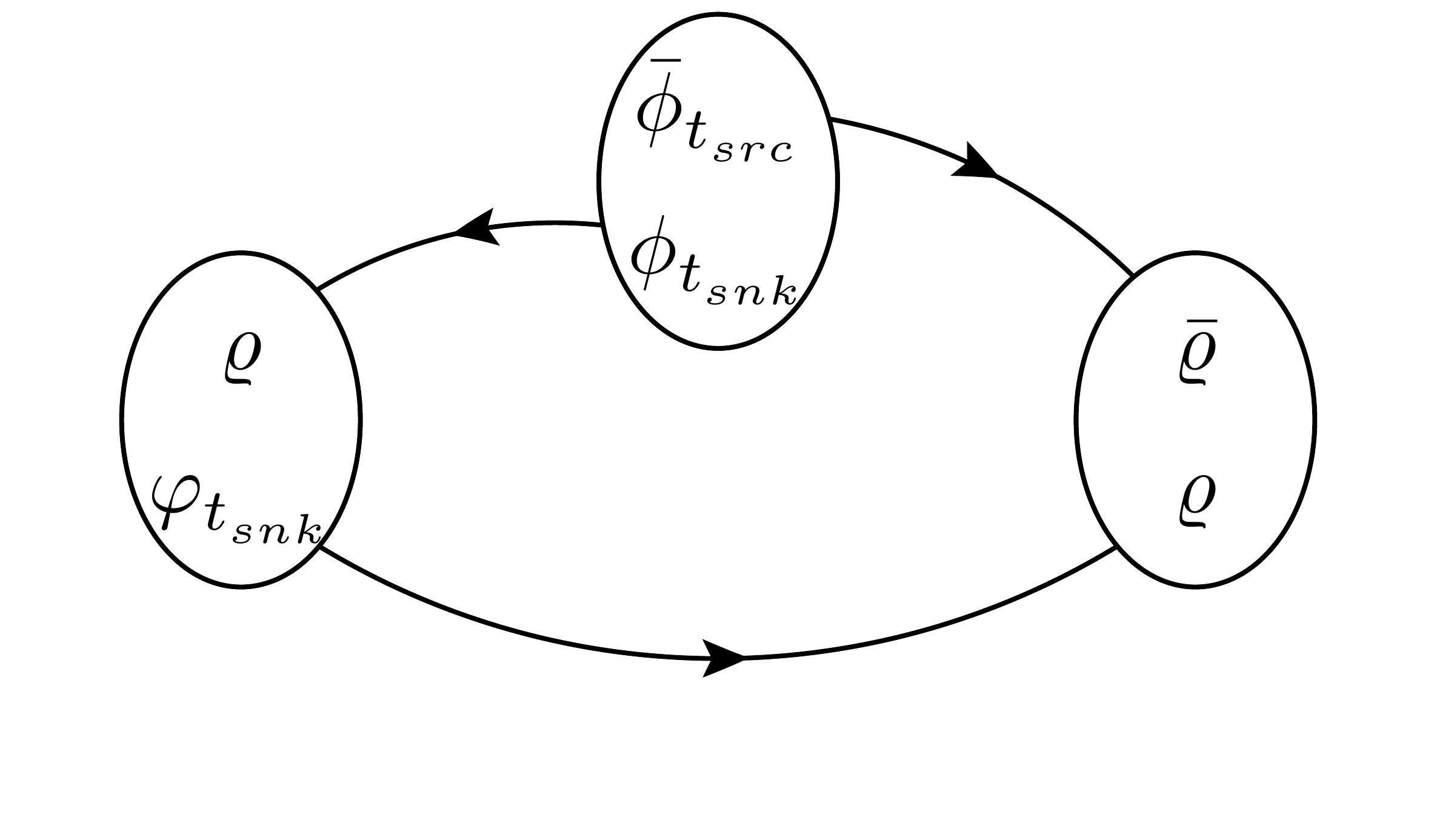}
  \caption{\textbf{Left panel:} Diagrammatic visualisation of the standard strategy to compute the three-point functions using a sequential solve on the $q_\mathrm{spec}$ propagator for each $\Gamma_\mathrm{snk}, \mathbf{p}_\mathrm{snk}, \Delta T$. \textbf{Right panel:} Diagrammatic sketch of $C_3$ evaluated using distillation. All operators $\Gamma$ as well as momenta $\mathbf{p}$ at all three positions as well as the distance $\Delta T$ can be chosen during the final (cheap) contraction.}
  \label{figure:hl-semi-z2}% Give a unique label
\end{figure}

A straightforward way to compute $C_3$ is to compute $D^{-1}_{q_\mathrm{spec}}$ first and then to sequentially invert on this propagator at $t_\mathrm{snk}$ with a $q_\mathrm{f}$ quark. This sequential inversion needs to be done for each $\Gamma_\mathrm{snk}, \mathbf{p}_\mathrm{snk}, \Delta T$. A sketch of this technique is shown in the left panel of Figure \ref{figure:hl-semi-z2}.
The choice of $\Delta T$ has to be a compromise, as the signal suffers from a bad signal-to-noise ratio for large $\Delta T$, but when choosing a smaller $\Delta T$ we cannot isolate the ground state - a suitable quark smearing technique might help to get a good ground state signal already for a smaller $\Delta T$.

One further difficulty is to obtain the overlap factor between a momentum-carrying meson and the vacuum, for which we need two-point functions with non-zero momentum built from a $Z_2$-wall source $\mathcal{Z}_2(x)$. To achieve this we create phased momentum sources
\begin{align}
\mathcal{Z}_2^\mathbf{p}(x) = \mathcal{Z}_2(x) \times e^{i\mathbf{p}\mathbf{x}}
\, .
\end{align}
A propagator computed from a phased source $\mathcal{Z}_2^\mathbf{p}(x)$ combined with a second propagator from the non-phased source $\mathcal{Z}_2(x)$ then gives the desired correlation function.

The idea of this work is to compare this traditional method to estimating $C_3$ using Distillation with LapH smearing \cite{Peardon:2009gh,Morningstar:2011ka}. This technique is based on a hermitian smearing matrix $S=\sum_{k=1}^{N_\mathrm{vec}} V_{k} V_{k}^\dagger$ constructed from $N_\mathrm{vec}$ low modes $V_{k}(x,t)$ of the 3D lattice Laplacian. Sources are created by applying dilution projectors \cite{Wilcox:1999ab,Foley:2005ac} onto these low modes, leading to the definitions
 \begin{align}
\varrho^{[d]}  = V P^{[d]} \eta\, , \phantom{aaa}
\phi^{[d]}  = D^{-1} \varrho^{[d]}\, , \phantom{aaa}
\tau^{[d]}  = V^\dagger \phi^{[d]}\, , \phantom{aaa}
\varphi^{[d]}  = V \tau^{[d]}
\, ,
\end{align}
with random noise vectors $\eta$\footnote{We use the notation developed for stochastic distillation throughout this paper, but we employ exact distillation which is restored by using full dilution and setting the noise vectors to $\rho=1$.}, dilution projectors $P^{[d]}$, the diluted LapH source vectors $\varrho^{[d]}$ and sink vectors $\varphi^{[d]}$, the unsmeared sinks $\phi^{[d]}$ \cite{Mastropas:2014fsa}, which can be used to define local currents, and the (stochastic) perambulators $\tau^{[d]}$ which are non-lattice sized (and therefore cheap to store) objects.

Using these definitions, the three-point function can be evaluated in the LapH framework by inserting the LapH-smearing matrices at the source and the sink, but not at the current insertion, which we require to be local. This can be straightforwardly evaluated to
 \begin{align}
\nonumber C_3 &= \langle \phi_{t_\mathrm{src}}^\dagger(t) \Gamma_\mathrm{op} \bar{\phi}_{t_\mathrm{snk}}(t) \varrho^\dagger(t_\mathrm{snk}) \Gamma_\mathrm{snk} \varphi_{t_\mathrm{src}}(t_\mathrm{snk}) \varrho^\dagger(t_\mathrm{src}) \Gamma_\mathrm{src} \bar{\varrho}(t_\mathrm{src}) \rangle \\
& = M_{\Gamma_\mathrm{op}}(\bar{\phi}_{t_\mathrm{snk}},\phi_{t_\mathrm{src}},t) M_{\Gamma_\mathrm{snk}}(\varphi_{t_\mathrm{src}},\varrho,t_\mathrm{snk}) M_{\Gamma_\mathrm{src}}(\bar{\varrho},\varrho,t_\mathrm{src})^*
\, ,
\label{eqn:3pt-dist}
\end{align}
\vspace{-1cm}
\begin{align}
M^{[d_1,d_2]}_\Gamma(\varphi,\varrho,t,\mathbf{p}) = \sum_{\mathbf{x}} e^{-i\mathbf{p} \cdot \mathbf{x}} \varphi^{[d_1]*}(t) \Gamma \varrho^{[d_2]}(t) 
\, ,
\label{eqn:mf}
\end{align}
where $ \bar{\varrho} = \gamma_5 \varrho , \bar{\varphi} = \gamma_5 \varphi, \bar{\phi} = \gamma_5 \phi$, which arise when using $\gamma_5$ hermiticity to evaluate the $q_\mathrm{i}$ propagator. Diagrammatically, Equation \ref{eqn:3pt-dist} is shown in the right panel of Figure \ref{figure:hl-semi-z2}.

\section{Pseudoscalar-axial diagonalisation}
We have computed all combinations of the two-point function with pseudoscalar/axial ($\gamma_5, \gamma_0 \gamma_5$) at source and sink and expect them to behave like
\begin{align}
C_{ij}(t) = \begin{pmatrix} C_{PP} & C_{PA} \\ C_{AP} & C_{AA} \end{pmatrix}
= \begin{pmatrix} \langle O_P O_P^\dag \rangle & \langle O_P O_A^\dag \rangle \\ \langle O_A O_P^\dag \rangle & \langle O_A O_A^\dag \rangle \end{pmatrix}(t) = \sum_{i=0}^1 \begin{pmatrix} A_{P_i}^2 & A_{A_i} A_{P_i} \\ A_{A_i} A_{P_i} & A_{A_i}^2 \end{pmatrix} e^{- E_i t}
\end{align}
where we have averaged the forward- and backward-propagating contribution and allow for the ground state and one excited state to be present\footnote{For the fit, we neglect the backward-propagating contribution, which only plays a role close to the centre of the lattice from which we stay away.}. We can perform a 6-parameter fit ($A_0$, $P_0$, $E_0$, $A_1$, $P_1$, $E_1$) to this equation and to extract matrix elements and energy levels. Furthermore, we define new operators $O'_P = \cos \theta O_P$ and $O'_A =\sin \theta O_A,$ with a tunable parameter $\theta$, leading to the correlation function
\begin{align}
C (\theta) &= \langle O' (\theta) O'^\dag (\theta) \rangle
= \cos^2 \theta C_{PP} + \cos \theta \sin \theta (C_{PA} + C_{AP}) + \sin^2 \theta C_{AA}
\, .
\label{eqn:c-theta}
\end{align}
By varying $\theta$ we can define several "diagonalised" operators, some of which may have earlier plateaus - due to cancellation of axial and pseudoscalar excited states - than the correlation functions directly computed from the lattice. Because $C (\theta)$ is built both from correlation functions with a $\sinh$ and $\cosh$ dependence on the mass, we restrict ourselves to $t<<T/2$ when studying them.

\section{Discussion of the data}

\begin{figure}[htbp] % no figure before 1st section
  \centering
\includegraphics[width=0.91\textwidth]{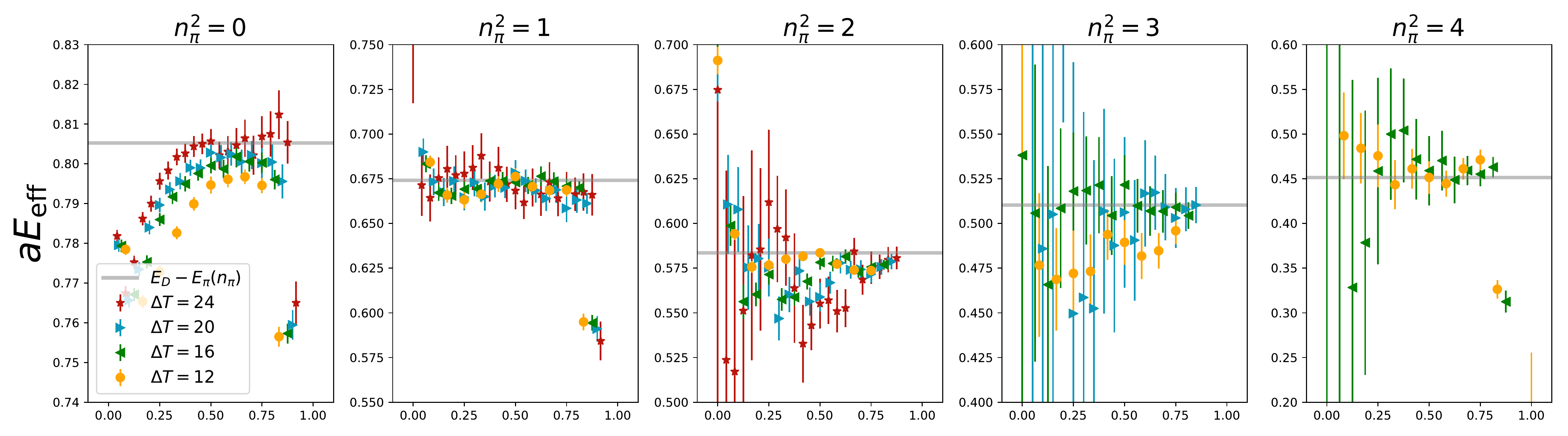}
\includegraphics[width=0.91\textwidth]{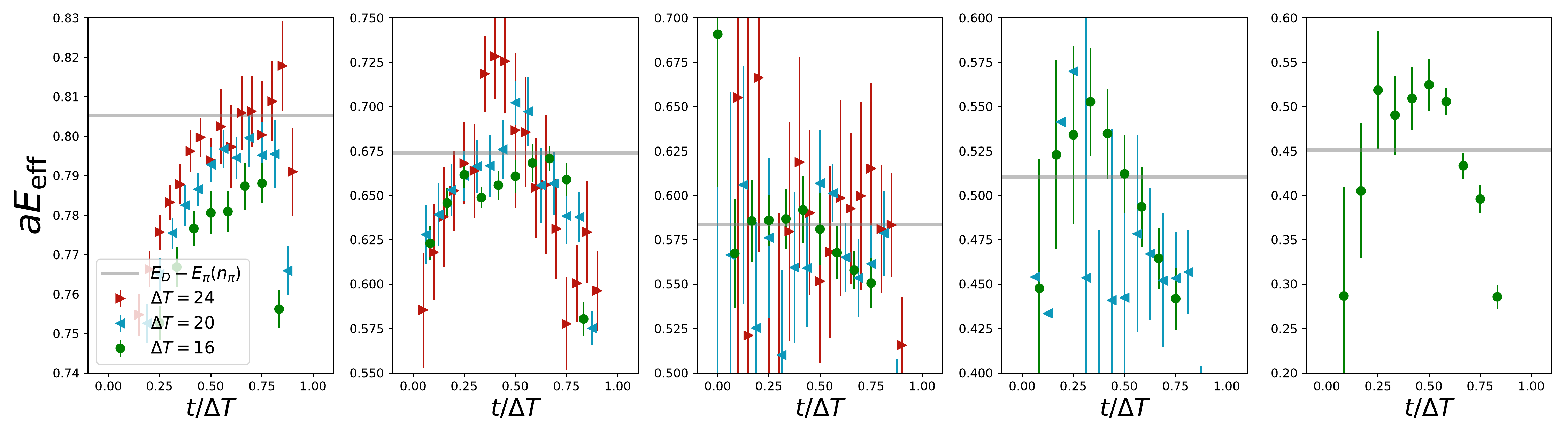}
  \caption{Comparison of the effective energies of the three-point functions for different values of $\Delta T$, showing the five (pion-)momentum frames $0\leq\mathbf{n}_\pi^2\leq4$ (where $\mathbf{p} = \frac{2 \pi}{L} \mathbf{n}$) and current insertion $\Gamma_\mathrm{op}=\gamma_0$ and $\Gamma_\mathrm{src} = \Gamma_\mathrm{snk} = \gamma_5$. The horizontal axis has been scaled to have the initial state ($D({\bf p}_D)$) at 0 and the final state ($\pi({\bf p}_\pi$)) at 1. The grey bands indicate the expected plateau obtained from $E_D - E_\pi(\mathbf{p}_\pi)$, which is computed using the lattice dispersion relation and from $am_D = 0.99656(95)$\cite{Boyle:2018knm}. The top panels show the distillation data and the bottom panels show the $Z_2$-wall data.}
  \label{figure:compare-delta}% Give a unique label
\end{figure}

\begin{figure}[htbp] % no figure before 1st section
  \centering
\includegraphics[width=0.9\textwidth]{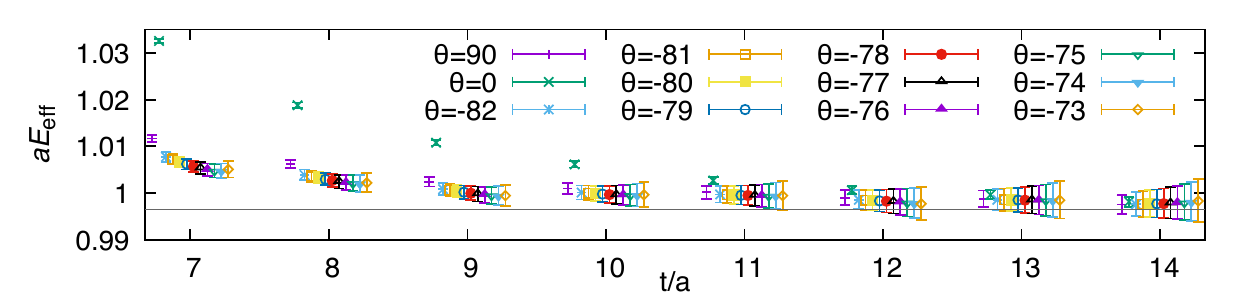}
  \caption{Effective energies of two-point correlation functions $C (\theta)$ defined in \ref{eqn:c-theta}. $\theta=0 \degree$ is proportional to $C_{PP}$ and $\theta=90 \degree$ is proportional to $C_{AA}$.}
  \label{figure:pa-diag}% Give a unique label
\end{figure}

We use a $64 \times 24^3$ RBC-UKQCD $(2+1)$-flavour domain-wall fermion gauge-field ensemble \cite{Allton:2008pn} with a pion mass of $m_\pi=329$ MeV and a lattice spacing of $a=0.11$ fm. We simulate $D \rightarrow \pi$ decay with a heavy mass $am_h=0.58$ and the same action as in \cite{Boyle:2018knm}. The current level of statistics for distillation is 5 configurations with 16 solves per flavour on each, averaged over the estimators with exchanged $q_\mathrm{i}$ and $q_\mathrm{f}$. We computed $4$ different $\Delta T \in \{12,16,20,24\}$. A comparison of the effective energy plateaux of the three-point functions for different $\Delta T$ is shown in Figure \ref{figure:compare-delta}; in the top panels for the distillation data and in the bottom panels for the $Z_2$-wall source data.
Figure \ref{figure:pa-diag} shows that by choosing $\theta=-79 \degree$ an earlier onset to the plateau can be achieved for the case of $Z_2$-wall sources. We cannot observe such an effect for the distillation data with the current level of statistics, possibly due to already suppressed excited states as a result of the LapH smearing.

\begin{figure}[htbp] % no figure before 1st section
  \centering
\includegraphics[width=0.91\textwidth]{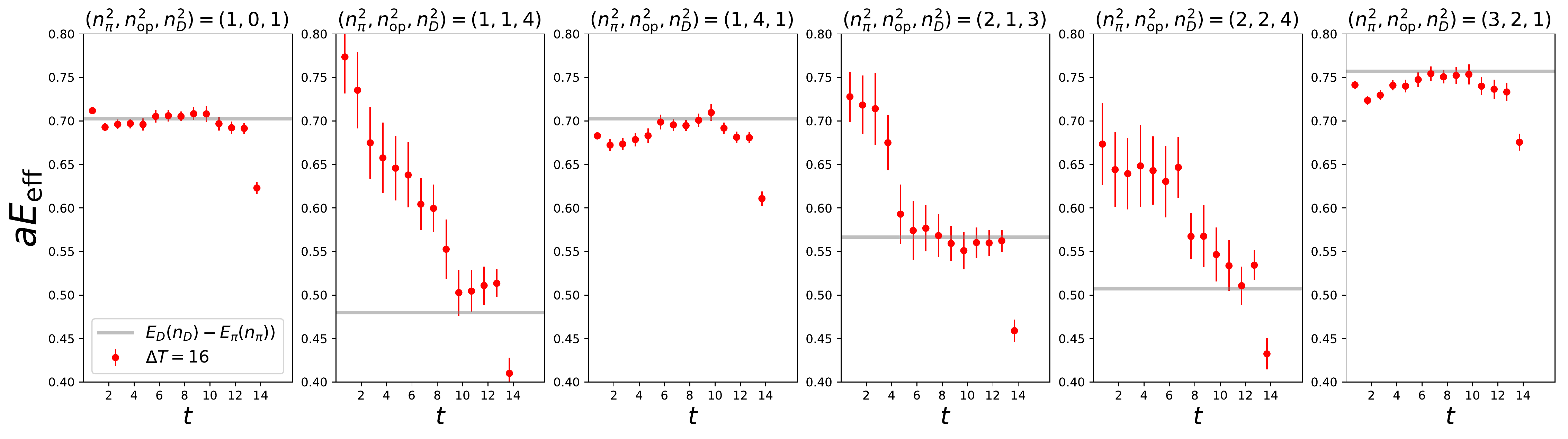}
  \caption{Effective energies of three-point correlation functions with non-zero momenta both at source and sink, only for $\Delta T = 16$, $\Gamma_\mathrm{src} = \Gamma_\mathrm{snk} = \gamma_5$ and $\Gamma_\mathrm{op}=\gamma_0$. The individual panel descriptions denote the 3-momenta involved, which obey $\mathbf{p}_\pi+\mathbf{p}_\mathrm{op}+\mathbf{p}_D =\mathbf{0}$. All momenta combinations are averaged over all possible lattice rotations. The grey bands indicate the expected plateau obtained from $E_D(\mathbf{p}_D) - E_\pi(\mathbf{p}_\pi)$, which is computed using the lattice dispersion relation and from $am_D = 0.99656(95)$\cite{Boyle:2018knm}.}
  \label{figure:diff-mom}% Give a unique label
\end{figure}

For the distillation data, we also show three-point correlation functions where both the mesons at source and sink have a momentum $\mathbf{p} \neq \mathbf{0}$. They can be cheaply assembled re-using already computed objects. We show our data for selected channels in Figure \ref{figure:diff-mom}.
Data extracted from these channels can be used to get additional energy levels and map out the $q^2$ region in between the values obtained for $\mathbf{p}_D=\mathbf{0}$. The quality of the effective energy plateaux varies a lot between channels and most are noisy compared with the plateaux of Figure \ref{figure:compare-delta}. In the $Z_2$-wall approach, the computation of these correlation functions would be costly and therefore impractical.

\section{Comparison of cost and statistics}

The computational cost of these calculations is summarised in Table \ref{tab:cost-small}.
\begin{table}
\centering
\begin{tabular}{|c|c|c|}
\hline 
  & Distillation & $Z_2$ seq. \\ 
\hline 
\#Inv / conf / $t_\mathrm{src}$ & $N_{vec} \times 4$ & $N_{\Delta T} \times N_{\mathbf{p}} \times N_{\Gamma_\mathrm{snk}}$ \\ 
\hline 
total \#Inv & 12800 & 1008 \\ 
\hline 
\end{tabular} 
\caption{Comparison of the number of inversions in the two approaches employed. This is just a rough measure for the computational cost, as the cost per inversion differs significantly for different quark flavours.}
\label{tab:cost-small}
\end{table}
The difference in the number of inversions for this setup amounts to a factor of about 13. When comparing the effective energies in all frames for the raw pseudoscalar and axial correlation functions, the error reduction is about a factor of $\approx 3-6$, i.e. no method is giving better statistical properties for the same computational cost\footnote{The reason why we get larger effective statistics using distillation is that we can compute all momentum rotations very cheaply and average over them, whereas in the $Z_2$-wall approach we compute just a single momentum direction per momentum frame.}. Different choices of $\Gamma_\mathrm{src}$ and $\Gamma_\mathrm{snk}$ do not come at additional cost for distillation. This allows the pseudoscalar-axial diagonalisation but it also allows for vector states at no additional cost. The inherent smearing when using distillation leads to slightly flatter plateaux, but a similar effect can be achieved by employing our suggested pseudoscalar-axial diagonalisation. 

Distillation could be more cost-effective by using a more ambitious setup, like $q_\mathrm{spec}=l,s$, $q_{f}=l,s,h$, $q_{i}=$ multiple heavy-quark masses, which would allow us to study the processes $D \rightarrow \pi$, $D \rightarrow K$, $D_s \rightarrow K$, $D_{(s)} \rightarrow D^{'}_{(s)}$. While the inversion cost for the distillation run would only increase linearly with the new $s$ and $h$ quarks, the amount of inversions in the $Z_2$-wall approach would increase with the number of combinations of $q_\mathrm{spec},q_\mathrm{f}$ quarks due to the sequential inversions. When using distillation, one would also need additional storage for the fields needed to assemble correlation functions. On our $24^3$ ensemble, the meson fields amount to $25$ TB per configuration. They can be deleted after all the contractions are completed. On a physical-pion-mass ensemble with a larger physical volume $V$, the storage needed will scale with $O(V^2)$, as $N_\mathrm{vec}$ would need to scale with $O(V)$, if the smearing radius is to be kept the same. Stochastic distillation would lessen this effect however by stochastically sampling the sources in the Laplacian-eigenvector space, so that the scaling in storage space needed would be mild.

\section{Conclusions and Outlook}

We have computed two-point and three-point correlation functions to study heavy-light semileptonic decays using two different approaches: Once by using $Z_2$-wall sources and sequential solves and once using distillation and have compared the cost of the two approaches\footnote{We note that there are further viable approaches like Coulomb gauge fixed approaches such as the gauge fixed wall approach, which are not part of this study.}. We find that both approaches have clear advantages, but that there is no approach which is better in all circumstances:

Distillation could be useful to study a very big project looking at many semileptonic decay channels. The method is very effective, but the cost both for the computation and intermediate storage needed cannot be neglected. This comes in particular from the local current insertion in the three-point functions. This problem would increase if we were to run this on the new RBC-UKQCD domain-wall ensemble at physical pion mass \cite{Blum:2014tka}, even though the use of stochastic distillation would lessen this problem to some degree. Distillation might also be the tool of choice if the plan of the study is to get results for many momentum transfers $q^2$. 

The traditional approach using $Z_2$-wall sources with sequential solves however is better suited for a smaller-scale project, where only a few decay channels are considered. It does not perform worse than distillation in this case, but is a lot easier to set up and does not need the large amount of intermediate storage for the meson fields. 

\textbf{Acknowledgements}
The  authors  thank  the members of  the  RBC  and  UKQCD  Collaborations  for
helpful discussions and suggestions. This work used the DiRACExtreme Scaling service at the University of Edinburgh, operated by the Edinburgh Parallel Computing Centre on behalf of the STFC DiRAC HPC Facility (www.dirac.ac.uk). The equipment was funded by BEIS capital funding via STFC grants ST/R00238X/1 and ST/S002537/1 and STFC DiRAC Operations grantST/R001006/1. DiRAC is part of the National e-Infrastructure. F.E. and A.P. are supported in part by UK STFC grant ST/P000630/1. F.E. and A.P. also received funding from the European Research Council (ERC) under the European Union's Horizon 2020 research and innovation programme under grant agreement No 757646 \& A.P. additionally by grant agreement 813942. J.T.T. is thankful for support by the Independent Research Fund Denmark, Research Project 1, grant number 8021-00122B. P.B. received support from the Royal Society Wolfson Research Merit award WM/60035.

\bibliographystyle{JHEP}
\bibliography{bib}

\end{document}